\documentclass[proof]{pasj00}
\draft

\begin{document}
\SetRunningHead{M. Nakagawa et al.}{Molecular Clouds in the Galactic Warp}
\Received{2005/01/13}
\Accepted{yyyy/mm/dd}

\title{An Unbiased Survey for Molecular Clouds in the Southern Galactic Warp}

\author{Masanori \textsc{Nakagawa},\altaffilmark{1}}
\email{masa@a.phys.nagoya-u.ac.jp}
\author{Toshikazu \textsc{Onishi},\altaffilmark{1} Akira \textsc{Mizuno},\altaffilmark{2} and Yasuo \textsc{Fukui}\altaffilmark{1}}
\altaffiltext{1}{Department of Astrophysics, Nagoya University, Nagoya 464-8602}
\altaffiltext{2}{Solar-Terrestrial Environment laboratory, Nagoya University, Toyokawa 442-8507}


\maketitle

\begin{abstract}
We have made an unbiased survey for molecular clouds in the Galactic
Warp. This survey, covering an area of 56 square degrees at $l$ =
252$\arcdeg$ to 266$\arcdeg$ and $b$ = $-$5$\arcdeg$ to $-$1$\arcdeg$, 
has revealed 70 molecular clouds, while only
6 clouds were previously known in the region. 
The number of molecular clouds is,
then, an order of magnitude greater than previously known in this sector 
at $R \gtrsim$ 14.5 kpc. The mass of the clouds
is in a range from 7.8$\times 10^{2} M_{\odot}$ to 8.4$\times 10^{4}
M_{\odot}$, significantly less than the most massive giant molecular 
clouds in the inner disk, $\sim$10$^{6} M_{\odot}$, while the cloud
mass spectrum characterized by a power law 
is basically similar to other parts of the Galaxy. 
The $X$ factor, $N$(H$_{2}$)/$\int T(^{12}{\rm CO}) dV$, 
derived from the molecular clouds in the Warp
is estimated to be 3.5($\pm$1.8) times larger than that in the inner disk. 
The total molecular mass in the Warp 
is estimated as 7.3$\times$10$^5$ $M_{\odot}$, and total mass 
in the far-outer Galaxy ($R \gtrsim$ 14.5 kpc) can be estimated 
as 2$\times$10$^7$ $M_{\odot}$.
The spatial 
correlation between the CO and H{\small \,I} distribution appears
fairly good, and the mass of the molecular gas is about 1\% of 
that of the atomic gas in the far-outer Galaxy.
This ratio is similar to that in the interarm 
but is ten times smaller than those of the spiral arms. 
Only 6 of the 70 Warp clouds show signs of star formation at the 
IRAS sensitivity and star formation efficiency 
for high-mass stars in the Warp is found to be smaller than 
those in other molecular clouds in the Galaxy.
\end{abstract}

\section{Introduction}
Molecular clouds are the sites of star formation, and play a crucial role in controlling the evolution of galaxies. It is therefore important to understand the physical conditions of molecular clouds and star formation therein under various physical conditions. 

Previous efforts to study molecular clouds have been mostly concentrated toward the inner Galaxy. These studies revealed the existence of giant molecular clouds (GMCs) having masses of 10$^5$ to 10$^6$ $M_{\odot}$, and that GMCs are actively forming both high-mass and low-mass stars (e.g., \cite{moo88, bli93}). The outer part of the Galactic disk is probably under significantly different physical conditions compared to the inner disk in several respects, including weaker stellar gravity, lower H{\small \,I} density, less intense UV fields and smaller cosmic ray flux (e.g., \cite{blo84}). It is also known that the metallicity becomes lower in the outer Galaxy (e.g., \cite{rud97}). Therefore, naturally we need to study the outer Galaxy, a unique nearby laboratory, to make new tests on how properties of star formation such as mass function, star formation efficiency etc. may change, depending on the above physical parameters.

Optical components in the outer Galaxy are being revealed gradually. A deep photometric survey of late-type stars (mostly G and K dwarfs) toward the Galactic anticenter by \citet{rob92} indicates a sharp cutoff of the old stellar disk beyond $R$ = 14 kpc.  On the other hand, Population I stars apparently have a greater radial extent although they are not so many. Some H{\small \,II} regions have been found and studied at the outer edge of the Galaxy at $R$ = 15-19 kpc, suggesting that star formation is taking place there at a lower level (e.g., \cite{vil96, deg93, kob00, san00}). It is thus very intriguing to study the dense molecular clouds and star formation in the outer Galaxy.

To reveal the molecular distribution in the outer Galaxy, several efforts have been made to date to search for molecular clouds (e.g., \cite{kutm81, wou89, sod91, bra94, dig94, may97}). \citet{kutm81} were the first to report the detection of low-level CO emission from molecular clouds in the first quadrant of the outer Galaxy. Although these studies cover a range of galactocentric distances ($R$) from 8.5 kpc to 20 kpc and distances from the Galactic plane ($|z|$) of up to 1.5 kpc, they are limited in spatial resolution \citep{may97, sod91} or spatial coverage \citep{dig94} or are biased toward bright IRAS point sources \citep{wou89, bra94}. \citet{wou89} observed $^{12}$CO($J$ = 1--0) emission with SEST and IRAM telescopes in the direction of 1302 cold IRAS sources in the second and third Galactic quadrants and gave a catalog of those clouds (WB clouds, hereafter), providing a most comprehensive view of star-forming molecular clouds in the outer Galaxy \citep{wou89, bra94}. We shall hereafter refer to the outer edge of the part, $R \gtrsim$ 14.5 kpc, as the far-outer Galaxy following \citet{bra95} who first used this word for $R \gtrsim$ 16 kpc. Nonetheless, it is still not well known if there is a significant amount of non-star-forming molecular gas in the outer Galaxy.  The importance of an unbiased study of molecular clouds covering non-star-forming clouds has been demonstrated by a number of authors over a large mass range, including those papers on cloud cores (e.g., \cite{oni96, oni98, oni02}) and on the giant molecular clouds in the Large Magellanic Cloud and M33 \citep{fuk99, eng03}.

Most recently, \citet{sne02} studied the properties of star-forming regions associated with 63 outer Galaxy ($R \geq$ 13.5 kpc) molecular clouds, among which the most massive ones have molecular masses of $\sim$10$^4$ $M_{\odot}$, within a 60 deg$^2$ area toward the second Galactic quadrant based on an unbiased CO Survey of the outer Galaxy at FCRAO \citep{hey98, hey01}. The survey has shown that the number of the molecular clouds dramatically decreases beyond $R$ = 13.5 kpc. 
About 10 clouds among the 63 have been found to be forming stellar clusters as revealed by K'-band imaging, and 39/246 clouds at $R \geq$ 13.5 kpc have 2MASS extended source(s) that are candidates for star-forming sites \citep{sne02}. 
\citet{bru03} decomposed the FCRAO data by applying a lower threshold for radiation temperature and doubled the sample of molecular clouds in the far-outer Galaxy in the second Galactic quadrant.
These 2 papers show a possibility that a significant fraction of molecular clouds quiet in star formation still remains unobserved in the far-outer Galaxy. It is therefore an important task to better and firmly establish the properties of both star-forming and non-star-forming molecular clouds in order to have a comprehensive understanding of the molecular gas in the outer Galaxy.  This goal should be attained by a large unbiased-sample of molecular clouds in the far-outer Galaxy. 

In this paper we describe the first results of the CO $J$ = 1--0 survey of the third Galactic quadrant covering 56 square degrees for a velocity range corresponding to the far-outer Galaxy at $R$ = 14.5-19 kpc. We shall here focus on the southern Warp which shows a large shift from $b$ = 0$\arcdeg$ by keeping in mind that the Warp is potentially an ideal place to avoid contamination from the Galactic disk at every wavelength. The present work should be complementary to the work by \citet{sne02} which studied the northern Warp. In this paper, details of the observations are given in section 2 and the results in section 3. Analysis and discussion are given in sections 4 and 5, respectively. Conclusions are given in section 6.

\section{Observations}
Observations were made in the $J$ = 1--0 $^{12}$CO emission line at 2.6 mm wavelength. The observed region is a part of the Warp in the outer edge of the Galaxy. The distribution of the Galactic Warp was first recognized in the 21 cm H{\small \,I} emission \citep{ker57} and \citet{bur86} revealed a pair of significantly warped regions in the H{\small \,I} disk up to 1.5 kpc in $|z|$. More recent papers on the H{\small \,I} Warp may be found elsewhere (e.g., \cite{gar02}). The present region is the southern part of the Warp, located mainly in the Galactic third quadrant. The Parkes 21 cm H{\small \,I} survey \citep{ker86} was used to select the region at $l$ = 252$\arcdeg$-266$\arcdeg$, and $b$ = $-$5$\arcdeg$-$-$1$\arcdeg$, where the H{\small \,I} emission is relatively strong at $R$ $>$ 14.5 kpc and the angular deviation of the Warp from $b$ = 0$\arcdeg$ is largest. The area observed corresponds to about 56 square degrees in the sky.

Observations were made with the NANTEN telescope at the Las Campanas Observatory in Chile in the period from July 2001 to August 2002. The main-dish diameter of the telescope is 4 m, providing a half-power beamwidth (HPBW) of 2.6 arcmin at 2.6 mm wavelength. This corresponds to a spatial resolution of 7.5 pc at a distance of 10 kpc, a typical distance to the southern Warp region. The front-end was an SIS mixer receiver. System temperature including the atmosphere was typically 200-250 K in SSB toward the zenith \citep{oga90}. We used two types of spectrometers. One of them was an acousto-optical spectrometer (AOS) which provided a velocity resolution of 0.65 km s$^{-1}$ with a velocity coverage of 650 km s$^{-1}$ (the Wide Band (WB) AOS). The other had a velocity resolution of 0.1 km s$^{-1}$ for a velocity span of 100 km s$^{-1}$ (the Narrow Band (NB) AOS). The pointing accuracy was better than 20 arcsec, as checked by optical observations of stars with a CCD camera attached to the telescope, as well as by radio observations of Jupiter, Venus, and the edge of the Sun. In reducing the spectral data, we subtracted second-order or third-order polynomials determined from the emission-free velocity parts in order to ensure a flat spectral baseline.

An absolute intensity calibration to obtain $T_{\rm A}^*$ was made in the chopper-wheel method. In order to convert $T_{\rm A}^*$ into $T_{\rm R}^*$ we assume that target sources are extended compared with beam size. We observed Orion KL [$\alpha$(1950) = 5$^{\rm h}$32$^{\rm m}$47.$^{\rm s}$0, $\delta$(1950) = $-$5$\arcdeg$24$\arcmin$21$\arcsec$] every two hours, and the peak $T_{\rm R}^*$ was taken as 65 K. All the observations were performed in the position-switching mode. ``Off'' positions, chosen to be as close as possible to ``On'' positions within two degrees, were checked to be emission-free at rms noise level of 0.07 K/ch in $T_{\rm R}^*$ for both (WB and NB) AOSs. First, we observed the whole region at a 4 arcmin grid spacing with WB AOS. The coverage is thus under-sampled with a sampling fraction of 33\%. The rms noise level in these observations was 0.2 K for 0.65 km s$^{-1}$ resolution. Next, we observed the regions of CO emission detected at a 2$\sigma$ noise level at $V_{\rm LSR}$ $\geq$ 90 km s$^{-1}$ at a 2 arcmin grid spacing with NB AOS. The rms noise fluctuations were 0.25 K for 0.1 km s$^{-1}$ resolution. If we adopt a flat rotation curve ($R_{\odot}$ = 8.5 kpc, $V_{\odot}$ = 220 km s$^{-1}$, \cite{kerl86}), the velocity threshold of 90 km s$^{-1}$ corresponds to $R$ = 14.9 kpc at $l$ = 252$\arcdeg$ and $R$ = 14.4 kpc at $l$ = 266$\arcdeg$, respectively, and the maximum velocity covered $V_{\rm LSR}$ = 160 km s$^{-1}$ corresponds to $R$ = 36 kpc at $l$ = 252$\arcdeg$ and $R$ = 31 kpc at $l$ = 266$\arcdeg$, respectively. The total numbers of the observed positions in the 4 arcmin and 2 arcmin observations were $\sim$13000 and $\sim$7500, respectively. We admit that the under-sampling in the first step observations places a lower limit for the detectable cloud mass. Nonetheless, as is shown later in section 3.1, we are confident that the present survey is complete in detecting molecular clouds whose mass is greater than 3000 $M_{\odot}$.

\section{Results}

\subsection{CO distribution and cloud catalog}
Figure 1 shows the velocity integrated intensity distribution of CO in the Warp at $V_{\rm LSR}$ $\geq$ 90 km s$^{-1}$. We have detected significant CO emission towards 702 positions at the 3$\sigma$ level, $\sim$0.9 K km s$^{-1}$ in CO integrated intensity, and have identified 70 individual molecular clouds at the same level (0.9 K km s$^{-1}$). The CO emitting area with the present beam is 0.78 deg$^2$, 1.4\% of the observed area, indicating that the molecular clouds are very sparsely distributed in the region. Among the 70 clouds, 64 are newly detected, corresponding to 91\% in number, and six of the 70 molecular clouds had already been detected and/or mapped by the previous observations. The total molecular mass of the 70 clouds is estimated to be 7.3$\times$10$^5$ $M_{\odot}$ (see section 5.2), as compared to $\sim$1$\times$10$^5$ $M_{\odot}$ previously known. In this work we have discovered molecular clouds an order of magnitude greater than previously known in this sector both in number and in mass, as explained in detail later in section 3.2. In Figure 2a we present some typical CO profiles which exhibit peak $T_{\rm R}^*$ of 1-2 K with a small linewidth of a few km s$^{-1}$. The present completeness limit in $L_{\rm CO}$ is estimated as $\sim$300 K km s$^{-1}$ pc$^2$, corresponding to $\sim$3000 $M_{\odot}$, which is set by the 4 arcmin grid and rms noise level of the first-step. On the other hand, in the second-step the present detection limit corresponds to $L_{\rm CO}$ $\sim$30 K km s$^{-1}$ pc$^2$ at $D$ = 10 kpc and $L_{\rm CO}$ $\sim$70 K km s$^{-1}$ pc$^2$ at $D$ = 15 kpc. The small cloud shown in Figure 2b is an example of a low mass cloud close to the detection limit in mass, $\sim$800 $M_{\odot}$. 

Table 1 lists the observed properties in CO and physical parameters of the 70 Warp clouds as follows; (1) cloud number, (2--3) peak position in Galactic coordinates, (4) central velocity with respect to the local standard of rest (a composite profile is fitted by a Gaussian curve to derive this quantity), (5--6) distance from the Sun as determined from the rotation curve and from the Galactic center projected at $b$ = 0$\arcdeg$ plane, (7) distance from the Galactic mid plane ($b$ = 0$\arcdeg$), (8) total integrated intensity, (9) full linewidth at half maximum (from Gaussian fit to a composite spectrum), (10) radius $r$ as derived from $r$ = ($S$/$\pi$)$^{1/2}$ where $S$ stands for the area at the three sigma intensity level, (11) beam-size corrected radius (here we assume that the beam pattern and cloud distributions are Gaussian, and cloud radii were corrected by using the following equation; (corrected size: $r_c$)$^2$ = (observed size: $r$)$^2$$-$(beam size)$^2$). (12) CO luminosity, $L_{\rm CO}$, calculated as (the total integrated intensity)$\times$(grid area), (13) mass, $M_{\rm CO}$, derived from $L_{\rm CO}$ by using the $X$ factor (see section 4 for details on the $X$ factor), (14) virial mass estimated as $M_{\rm vir}$ = 210$\times$ $r_{c}$(size [pc])$\times$ d$V^2$(linewidth [km s$^{-1}$]) by assuming that the cloud is a uniform sphere of radius (beam-size corrected radius: $r_c$), and (15) comments; ``cold IRAS'' indicates clouds which are associated with a cold IRAS point source (selection criteria; $F_{25} < F_{60}$ with qualities $\geq$ 2 at wavelengths of 25 and 60 micron) near the peak. 

\subsection{Comparison with the previous works}
We shall compare the present CO results with the previous works. \citet{bra94}, BW94 hereafter, selected and mapped twenty-seven WB clouds at $R \gtrsim$ 16 kpc, including thirty star-forming IRAS point sources. Among those, \citet{wou89} detected 6 molecular clouds at $V_{\rm LSR}$ $\geq$ 90 km s$^{-1}$ in the present observed region (WB1096, 1103, 1121, 1126, 1152, 1153) and BW94 mapped four of them in CO $J$ = 1--0 emission. The present work has detected four of the 6 clouds and these coincide with the WB clouds, respectively, as follows; No. 21 (WB1096), No. 27 (WB1126), No. 63 (WB1152) and No. 64 (WB1153). No. 71 (WB1077), located on the border of the present area, has not been fully mapped. The present table does not include the remaining two clouds (WB1103 and WB1121). WB1103 was mapped by BW94 and its $L_{\rm CO}$ is calculated to be 5.1 K km s$^{-1}$ pc$^2$, below the present detection limit. WB1121 was observed only toward one point, the IRAS source position, by \citet{wou89} and was not mapped by BW94. The single point CO spectrum gives a lower limit of $L_{\rm CO}$ as $\sim$30 K km s$^{-1}$ pc$^2$, somewhat smaller than the present detection limit, and WB1121 is perhaps missed because of its low $L_{\rm CO}$.

Our observations do not spatially resolve small structures of less than $\sim$10pc. Toward four molecular clouds No. 21, 27, 63 and 64, \citet{wou89} made CO observations with the 15 m SEST telescope and gave peak $T_{\rm A}^*$, (roughly equal to $T_{\rm R}^*$,) equal to 2.2-4.5 K, which is a few times higher than $T_{\rm R}^*$ we obtained. This indicates that the small-scale distribution within the clouds is beam-diluted. On the other hand, a lower resolution study at 8.8 arc-min beam by \citet{may97} detected only one cloud in this area at $l$ = 257.5$\arcdeg$, having 1.5$\times$10$^5$ $M_{\odot}$, corresponding a present cloud complex resolved into three clouds, No. 27, 28 and 30, with the 2.6 arcmin beam. 

\subsection{Mass spectrum}
The cloud mass, $M_{\rm CO}$, ranges from 7.8$\times$10$^2$ $M_{\odot}$ to 8.4$\times$10$^4$ $M_{\odot}$, and the largest mass is significantly smaller than that of giant molecular clouds in the inner disk, i.e., $\sim$10$^{6}$ $M_{\odot}$.
 
The mass distribution of 68 clouds out of 70 is shown in Figure 3. Two clouds (No. 18 and 56) are not included here because their masses are not well determined due to double peaked profiles (see also Table 1). 
The mass spectrum of the 47 clouds is well fitted by a single power law for a mass greater than 3000 $M_{\odot}$ by taking the maximum-likelihood method \citep{cra70}, N($\geq$ $M_{\rm CO}$/$M_{\odot}$) = 1.3$\times$10$^4$ $M_{\rm CO}^{-\alpha}$$-$5.8, where $\alpha$ = $-$0.68$\pm$0.16. The mass spectra studied for the Galactic CO clouds are well fitted by a power law with an index value of $\alpha$ $\sim$0.5-0.8 (e.g., \cite{hey01, yon97, dig96, bra95, sol89}). The present result is consistent with these studies within the error. 
This leads to the conclusion that the cloud mass spectrum characterized by a power law is basically the same throughout the Galaxy.
The slope of the mass spectrum for clouds with $<$ 3000 $M_{\odot}$ becomes shallower, indicating that the sampling is incomplete below this mass.

\subsection{Overall uncertainties of the cloud mass}
Individual Warp clouds show $V_{\rm LSR}$ around 100 km s$^{-1}$, and their galactocentric distances are estimated to be $R$ = 14.4-18.6 kpc. The distance given in Table 1 may bear errors by the following two reasons, the uncertainties in the rotation curve and the cloud-to-cloud random motion.  An assumption of a flat rotation curve may result in error up to about 15\% in distance (see e.g., \cite{bra93}). We note here that the rotation curve in the far-outer Galaxy is only poorly determined at present and that a flat rotation curve is fairly commonly observed in the outskirts of external spiral galaxies (see for a review e.g., \cite{sof01}). We also note that the present targeted region is less affected by uncertainties in kinematic distance compared to the regions close to the Galactic center or to the anti-center where the rotation curve affects the distance estimate more significantly. The other source of error is the random motion of the clouds.  If we adopt 5 km s$^{-1}$ as the typical cloud-to-cloud velocity dispersion as in the inner disk (e.g., \cite{isr84}), the distance ambiguity may be up to $\sim$8\% of those given in Table 1. 
Another possibility for uncertainties in distance is systematic large non-circular motions such as a streaming. This is difficult to estimate quantitatively, and is assumed here to be negligible because streaming is usually prominent only in the vicinity of spiral arms.
Thus, the overall uncertainty is up to 17\% in distance. Another source of uncertainty is calibration of the absolute flux of CO, about 10\%. By including all these, the total uncertainty in mass is estimated to be $\leq$ 38\% for a given $X$ factor.

\section{Derivation of the $X$ factor}
We shall estimate the $X$ factor (the conversion factor), $X$ = $N$(H$_{2}$)/$\int T(^{12}{\rm CO}) dV$ [cm$^{-2}$/K km s$^{-1}$], in the Warp. This is practically a useful factor to convert the CO luminosity to molecular mass and is also a fundamental factor that may reflect the physical properties of molecular clouds (e.g., \cite{pag01, str88}). The $X$ factor for the Galactic disk clouds is around 2$\times$10$^{20}$ [cm$^{-2}$/K km s$^{-1}$]. The $X$ factor is also written as $X$ = $M_{\rm CO}$/$L_{\rm CO}$ [$M_{\odot}$/K km s$^{-1}$ pc$^2$], e.g., $M_{\rm CO}$ = 4.3$\times$$L_{\rm CO}$ (if $X$ = 2$\times$10$^{20}$ [cm$^{-2}$/K km s$^{-1}$]).

We show an $L_{\rm CO}$-$M_{\rm vir}$ diagram in Figure 4. We have plotted 54 Warp clouds that have more than three consecutive observed points in order to reduce the effects of beam dilution in cloud size. In addition, we have included, for comparison, 216 Carina arm clouds (see section 5.1) observed with the same instrument (Matsunaga et al. 2005, in preparation) and 75 GMCs in the inner disk with known distances (see Table 1 in \cite{sol87} and Appendix). In this figure, we used size-corrected clouds (see section 3.1) both for the Warp and the Carina regions. All virial masses are derived from the calculation $M_{\rm vir}$ = 210$\times$$r$$\times$d$V^2$ by assuming for simplicity that density distribution is uniform in a cloud. The Warp clouds are indicated by crosses, those in the Carina arm by open circles and those in the inner disk by diamonds. For all the three regions, $L_{\rm CO}$ and $M_{\rm vir}$ show a good positive correlation with a slope of $\sim$1. This figure also indicates that $L_{\rm CO}$ is significantly different among the three regions for a given $M_{\rm vir}$. The Warp clouds generally show $L_{\rm CO}$ a factor of two smaller than that of the Carina clouds and a factor of three smaller than that of the inner disk clouds; these differences may depend on the range in $M_{\rm vir}$. It is to be noted that the slopes of the $L_{\rm CO}$-$M_{\rm vir}$ relation for the three regions are slightly different, 1.05($\pm$0.13) for the Warp region, 0.93($\pm$0.04) for the Carina region and 0.78($\pm$0.03) for the inner disk.

We shall estimate the difference in the $X$ factor by comparing the offset in $L_{\rm CO}$ for a given virial mass. We restrict ourselves to the mass range 10$^{4.0}$-10$^{5.1}$ $M_{\odot}$ by keeping in mind that the accuracy of mass for smaller clouds whose $M_{\rm vir}$ is less than 10$^4$ $M_{\odot}$ is worse than clouds whose $M_{\rm vir}$ is greater than 10$^4$ $M_{\odot}$. In this range, there are still a significant number of sample clouds (see Figure 4), i.e., 33 Warp clouds and 20 inner disk clouds, allowing us to make a reasonable comparison with statistical significance. 
The solid lines in Figure 4 indicate regression lines with a slope of 1.0 for the Warp and the inner disk clouds; they are expressed as log($M_{\rm vir}$) = log($L_{\rm CO}$)+1.41($\pm$0.25) and log($M_{\rm vir}$) = log($L_{\rm CO}$)+0.86($\pm$0.24), for the Warp and the inner disk, respectively. The averaged value of $L_{\rm CO}$ is then estimated to be a factor 3.5$\pm$1.8(1 $\sigma$) smaller than that for the inner disk.

In this analysis we did not make a correction for size for the inner clouds, nor take into account the possible overestimation of linewidth due to cloud overlap in the line-of-sight for all regions. If we consider these effects, the virial masses of inner clouds would decrease, resulting in a larger increase in the $X$ factor of the Warp. 

Gamma-ray observations have been used to estimate the $X$ factor in the inner disk, $X$ = 1.6$\times$10$^{20}$ [cm$^{-2}$/K km s$^{-1}$] \citep{hun97} and $X$ = 1.9$\times$10$^{20}$ [cm$^{-2}$/K km s$^{-1}$] \citep{str96}. We shall adopt here the former value of the $X$ for the inner disk, 1.6$\times$10$^{20}$ [cm$^{-2}$/K km s$^{-1}$], and the above estimate of the difference in $L_{\rm CO}$, 3.5. By taking these values we obtain the present $X$ = 5.6$\times$10$^{20}$ [cm$^{-2}$/K km s$^{-1}$] in the Warp.

It is suggested that the $X$ factor in the outer Galaxy (9 kpc $\lesssim R \lesssim$ 14 kpc) is a few times larger than in the inner Galaxy \citep{mea88, sod91, dig96}. 
The present $X$ factor at $R \gtrsim$ 14.5 kpc is consistent with these, indicating that $X$ factor becomes larger by a factor of 2-4 at $R \gtrsim$ 9 kpc. 
The difference of the $X$ factor may be understood as due to lower temperature (see next Sect. 5.1) and/or to lower metallicity in the Warp. In this paper, we calculate cloud masses by using $X$ = 5.6$\times$10$^{20}$ [cm$^{-2}$/K km s$^{-1}$] or the relationship of $M_{\rm CO}$ [$M_{\odot}$] = 12.1$\times$$L_{\rm CO}$ [K km s$^{-1}$ pc$^2$]. 

The present derivation of the $X$ factor is based on the assumption that virial mass ($M_{\rm vir}$) remains the same for a cloud of given mass. Almost all the CO clouds in the Warp have larger $M_{\rm vir}$ compared with $M_{\rm CO}$ and the ratio of $M_{\rm vir}$/$M_{\rm CO}$ ranges from 0.5 to 6.4 with an average value of 2.1 (Table 1).  It is possible that the clouds are in dynamical equilibrium under the external pressure or that they are transient features with a timescale of the order of the typical crossing time, 3$\times$10$^6$ yrs. Smaller clouds with $L_{\rm CO}$ $<$ 3$\times$10$^2$ K km s$^{-1}$ pc$^2$, in particular, may not be bound by self-gravity \citep{hey01}. The virial mass range used in the present work $\geq$ 10$^4$ $M_{\odot}$ perhaps guarantees the above assumption.

The present result on the $X$ factor is consistent with previous work by \citet{mea88, sod91, dig96} for the outer Galaxy.
On the other hand, \citet{bra95}, BW95, and \citet{sne02} derived a conclusion that the $X$ factor is not significantly different between the far-outer Galaxy and the inner disk. These authors used $^{13}$CO emission to estimate the molecular hydrogen column density by using an LTE analysis with an H$_{2}$/$^{13}$CO ratio similar to that in the inner disk \citep{sne02} and with a higher H$_{2}$/$^{13}$CO ratio extrapolated from $^{12}$CO integrated intensity \citep{bra95}. They argued that the thus derived molecular column density agrees with that derived from $^{12}$CO integrated intensity using the conventional $X$ factor of the inner disk, $\sim$2$\times$10$^{20}$ [cm$^{-2}$/K km s$^{-1}$], and concluded that the $X$ factor is uniform in the Galaxy. 
This argument however may not be justified because the difference in the density regime probed in $^{12}$CO and $^{13}$CO is not properly taken into account. It is well known that the $^{12}$CO emitting region is larger than the $^{13}$CO emitting region in a given cloud; a typical mass ratio of $^{12}$CO and $^{13}$CO emitting regions are $\sim$3 in the local clouds (see e.g., \cite{miz98, miza01, oni99, tac01, har99}), and this is generally interpreted in terms of different self-shielding against far ultraviolet radiation (e.g., \cite{war96}). In the velocity space, this difference manifests itself as the broader line profile in $^{12}$CO than in $^{13}$CO, and the $^{12}$CO line profiles indeed exhibit emission in a velocity range where no $^{13}$CO emission is seen. 
Taking this into consideration, we need to accept that the molecular column density derived from $^{13}$CO refers to only the $^{13}$CO emitting region, although the whole $^{12}$CO emitting region should be more extended and more massive than the $^{13}$CO emitting region. 
Therefore, the molecular column density derived from $^{13}$CO should be smaller than that derived from $^{12}$CO reflecting the $^{12}$CO/$^{13}$CO cloud mass ratio, $\sim$3; it is likely that the exact ratio in column density may depend on the cloud geometry. 
In order to reconcile this with the results of BW95 and \citet{sne02}, we have two alternatives; one is that the ratio H$_{2}$/$^{13}$CO may be smaller than BW95 assumed and the other is that the $X$ factor is larger in the Warp than in the inner disk. The latter is actually in accord with the present result and is perhaps more likely (see also Sect. 5.1). We thus confirm the present conclusion that the $X$ factor in the Warp is a factor of 3.5 larger than in the inner disk, noting that the conclusion by BW95 and \citet{sne02} needs to be carefully reexamined by considering the $^{12}$CO/$^{13}$CO cloud mass ratio.

\section{Discussion}

\subsection{Comparison of the Warp clouds with the Carina-Arm clouds}
In order to compare the Warp clouds with the clouds in the Galactic disk, we have created a diagram of physical parameters that include clouds in the Warp and the Carina arm as shown in Figures 5. We have chosen here the Carina arm because the region was observed with the same instrument, NANTEN (Matsunaga et al. 2005, in preparation), and because cloud parameters can be better defined with less confusion than in the inner disk. The CO data of Carina were taken at rms sensitivity $\sim$0.3 K with WB AOS, 3 times worse than that for the Warp clouds, and at a 4 arcmin grid spacing, twice of that for the Warp. These observational parameters place the lower limit in cloud mass 20 times higher than in the Warp.  We have thus used only the reliable sample clouds in Carina whose mass is greater than 16,000 $M_{\odot}$  and restricted the cloud distances from the Sun ($D$) in a range from 10.3 kpc to 15.3 kpc, the same range for the Warp clouds.

Figure 5 gives the size-linewidth relation for the same sample of CO, a relationship often used to characterize the physical properties of clouds. In this figure, we also plotted the beam-corrected size (see Sect. 3.1) of the Warp clouds (crosses) and the Carina arm clouds (circles) in addition to the far-outer Galaxy clouds in the second Galactic quadrant observed by \citet{hey01}. The broken line and dotted line are least-squares fits calculated by BW95 for the inner- ($R$ $<$ 8.5 kpc) and outer-Galaxy clouds ($R$ $>$ 8.5 kpc), respectively. The diagram shows that molecular clouds in the far-outer Galaxy, from NANTEN and FCRAO data, are roughly consistent with the outer Galaxy line by BW95. This also clearly indicates that the outer Galaxy clouds are characterized by linewidths a factor of two smaller than the inner Galaxy clouds (see also Figure 10). On the other hand, the Carina clouds are located between the two lines for the outer Galaxy and for the inner Galaxy. If correct, this suggests that the clouds in the disk may have a gradient in linewidth as a function of $R$, which should be tested by a more extensive dataset in the future.

Here, we will again discuss on the $X$ factor based on the above relations in order to obtain a deeper insight into the cause of the larger $X$ factor in the Warp. If we denote the cloud size in the far-outer Galaxy and in the inner disk as $r^{OUT}$ and $r^{IN}$, respectively, the size-linewidth relation in Figure 5 is expressed as d$V=k/$2$\times$($r^{OUT}$)$^{1/2}$ and d$V=k\times$($r^{IN}$)$^{1/2}$, respectively. On the other hand, BW95 (Fig. 13) showed that $L_{\rm CO} \propto$ d$V^4$; we have confirmed this relation holds also for the present Warp clouds(see Appendix and Figure 10). By using the above equations and $M_{\rm vir} \propto r\times$d$V^2$, we find that $L_{\rm CO}$ in the far-outer Galaxy is one fourth of $L_{\rm CO}$ in the inner disk. This offers a simple explanation of how the $X$ factor becomes larger by a factor of 3.5($\sim$4) in the Warp, providing further support for the present $X$ factor.

\subsection{Comparison with H{\small \,I}}
The CO distribution is superposed on the H{\small \,I} \citep{ker86} distribution in the $l$-$b$ diagram (Figure 6). This figure indicates that the H{\small \,I} and CO are well correlated with each other in the sense that the CO clouds, while much clumpier than H{\small \,I}, are distributed mostly toward the intense H{\small \,I} emission. To be more quantitative, 93\% (= 65/70) of the CO clouds are located within the 800 K km s$^{-1}$ contour level of H{\small \,I} in Figure 6.

Figure 7 shows the distance $z$ from the Galactic mid plane as a function of $R$. The peak positions of CO clouds are superimposed on an H{\small \,I} intensity map integrated between $l$ = 252$\arcdeg$ and 266$\arcdeg$ from the H{\small \,I} data. This clearly shows that the CO clouds are distributed up to $R$ = 18.6 kpc and there is no molecular cloud beyond this radius as is consistent with BW94. The trend that $|z|$ increases with $R$ is clearly seen, an obvious sign of the Warp. \citet{wou90} also showed the sign of warp structure by CO and H{\small \,I} observations on an $l$-$z$ map. Comparisons between the H{\small \,I} and CO in Figures 6 and 7 indicate that the Warp clouds are not distributed uniformly, but that they are distributed in some groups extended over a few degrees, coinciding with larger H{\small \,I} distributions of similar extents. In Figure 7 one also sees that the $z$-distribution of the molecular clouds becomes wider with increasing $R$, a sign of the flaring of the molecular disk.

Figure 8 shows a face-on view of the outer disk. The peak positions of CO clouds in the Warp and the Carina arm are superimposed on an H{\small \,I} column density map from the H{\small \,I} data at $V_{\rm LSR}$ $\geq$ 10 km s$^{-1}$ \citep{ker86}. We used an intensity-mass conversion factor of 1.82$\times$10$^{18}$ [cm$^{-2}$/K km s$^{-1}$] to calculate H{\small \,I} mass. Thin broken contours indicate a column density of 0.5$\times$10$^{21}$ cm$^{-2}$. For the Warp and the Carina arm, the total molecular mass is estimated as 7.3$\times$10$^5$ $M_{\odot}$ and 2.1$\times$10$^7$ $M_{\odot}$, respectively, while that of H{\small \,I} is calculated to be 4.4$\times$10$^7$ $M_{\odot}$ and 14.5$\times$10$^7$ $M_{\odot}$, respectively. We applied a standard conversion factor, $X$ = 2$\times$10$^{20}$ [cm$^{-2}$/K km s$^{-1}$], for the Carina clouds to calculate their masses. By using the total area (11 kpc$^2$ for the Warp and 24 kpc$^2$ for the Carina arm), averaged surface densities of H$_2$ and H{\small \,I} in the Warp and the Carina arm are calculated within the thick broken line and are listed in Table 2 along with the corresponding values in other regions of the Galaxy, i.e., the inner disk, the Perseus arm and the interarm, and in the LMC for reference.

We find that the molecular cloud mass in the Warp is $\sim$1\% of the total hydrogen mass. This ratio is significantly less than that in the Carina arm (14\%) or in the Perseus arm (11\%) or in the inner disk (50\%), but is similar to that in the interarm (1\%). The ratio is also much less than that in one of the nearest galaxies, the LMC \citep{miz01}. This result is consistent with the conclusion reached by \citet{sne02} for the northern far-outer Galaxy and suggests that the conversion efficiency from atomic to molecular gas is ten times smaller in the far-outer Galaxy than in the inner disk. The disk layer thickness (FWHM) of H{\small \,I} gas increases with $R$, 400 pc at $R$ = 10 kpc and 600 pc at $R$ = 16 kpc \citep{wou90}, and the H{\small \,I} mass density is calculated as 1.5$\times$10$^7$ $M_{\odot}$/kpc$^3$ and 0.7$\times$10$^7$ $M_{\odot}$/kpc$^3$ for the Carina arm and the Warp, respectively. We suggest that this H{\small \,I} density decrease by a factor of 2 may lead to the lower efficiency of conversion into molecular gas in the Warp.

We shall discuss the total molecular mass in the far-outer Galaxy. The total molecular mass is estimated to be 7.3$\times$10$^5$ $M_{\odot}$ in the present region of the southern Warp. The area of the present survey corresponds to 3.5\% of the area at $R$ = 14.5-19 kpc. If we assume the proportion of molecular cloud in the far-outer Galaxy is uniform, the total molecular mass in the far-outer Galaxy is estimated to be 2.1$\times$10$^7$ $M_{\odot}$ by simply multiplying 100/3.5 to 7.3$\times$10$^5$ $M_{\odot}$. 
We can also check the total molecular mass in the same way by using the FCRAO data from the on-line catalog \citep{hey01, bru03}. \citet{hey01} found 117 clouds at $R$ $\geq$ 14.5 kpc within a region extending 40 deg in longitude, three times larger than the area in our survey, and gave $L_{\rm CO}$ for each. The area covered corresponds to $\sim$6\% of the far-outer Galaxy. The total molecular mass in that area is calculated to be 5.0$\times$10$^5$ $M_{\odot}$ (total CO luminosity is 4.1$\times$10$^4$ K km s$^{-1}$ pc$^2$), if the same $X$ factor as in the Warp is applied. By assuming a uniform molecular cloud distribution over the far-outer Galaxy, the total molecular mass is estimated to be 8$\times$10$^6$ $M_{\odot}$, somewhat smaller than the above for the southern Warp. This may be due to a moderate asymmetry in molecular mass between the south and north, while the difference is only within a factor of two. We thus infer that the current best estimate for the total molecular mass in the far-outer Galaxy is 1-2$\times$10$^7$ $M_{\odot}$. The total H{\small \,I} mass in the far-outer Galaxy is $\sim$2$\times$10$^9$ $M_{\odot}$ as derived by a few groups \citep{wou90, wol03}, and the overall molecular/atomic ratio is 0.5-1\%. On the other hand, \citet{bro00} estimated the total molecular mass at 8.5 $< R <$ 14.5 to be 2$\times$10$^8$ $M_{\odot}$ ($X$ = 1.56$\times$ 10$^{20}$ [cm$^{-2}$/K km s$^{-1}$]). We thus infer that the total molecular mass in the far-outer Galaxy is 5-10\% (depending on $X$; see \cite{bro00}) of that in the outer Galaxy.

\subsection{Star formation in the Warp clouds}
In order to investigate star formation in the Warp clouds we have used the IRAS point source catalog to select possible candidates for YSOs associated with the 70 Warp clouds. Cold IRAS point sources ($F_{25}$ $<$ $F_{60}$ with qualities $\geq$ 2 at both bands 25 $\mu$m and 60 $\mu$m) are associated with the peak CO emission in only 6 clouds (see Figure 1). These clouds may be forming massive to intermediate mass (B type) stars as judged from their IRAS flux density; we roughly estimate the total far-infrared luminosity of these IRAS sources from the flux densities at 12-100 $\mu$m, as $L_{FIR}$ = 2-19$\times$10$^3$ $L_{\odot}$, where we use a simple method of estimating $L_{FIR}$ from the sum of the fluxes at 4 bands \citep{cas86}. 

The $L_{FIR}$/$M_{\rm CO}$ ratio for these 6 clouds ranges from 0.1 $L_{\odot}$/$M_{\odot}$ to 2.7 $L_{\odot}$/$M_{\odot}$ with an average value of 0.8 $L_{\odot}$/$M_{\odot}$. We note that this ratio is smaller than the previous studies. \citet{sne02} estimated the $L$/$M$ ratios, 0.4-33 $L_{\odot}$/$M_{\odot}$ with an average value of 8.1 $L_{\odot}$/$M_{\odot}$, based on 9 clouds in the far-outer Galaxy at $R$ = 13.5-17.4 kpc, where they used an $X$ factor of 1.9$\times$10$^{20}$, a factor of 2.9 smaller than the present value. If the same $X$ factor for the Warp, $X$ = 5.6$\times$10$^{20}$ [cm$^{-2}$/K km s$^{-1}$], is applied, averaged $L$/$M$ decreases to be 2.8 $L_{\odot}$/$M_{\odot}$. \citet{sne02} suggested that the $L$/$M$ ratios for the far-outer Galaxy clouds are similar to those for the inner disk clouds, which are in the range 0.5-30 $L_{\odot}$/$M_{\odot}$ with an average value of 4 $L_{\odot}$/$M_{\odot}$ \citep{moo88}. The present sample indicates that the $L$/$M$ ratio in the southern Warp is a factor of 4-5 smaller than that in the northern far-outer Galaxy and inner Galaxy. We therefore suggest that the star formation efficiency in the southern Warp may tend to be smaller although we definitely need to pursue this issue by increasing the number of sample clouds with more sensitive infrared source list in the far-outer Galaxy.

We have created a histogram of cloud mass. Figure 9 indicates that massive clouds having larger molecular column density form stars (dark gray) more often than lower mass ones, a similar trend to the disk cloud (e.g., \cite{yon97, kaw98, har99, tac02, oni02}). The trend seems to be consistent but not so strong as in the case of the disk cloud core, possibly due to different range of cloud mass, the lower spatial resolution of the present molecular data and the lower sensitivity of the infrared sources. We suggest that the present molecular clouds provide a set of good targets which can be used to make detailed searches for star formation at various wavelengths, in particular at near- to mid-infrared wavelengths.

It has been shown by several authors that the ratio between the virial mass and the luminous mass of molecular clouds is a good indicator of the gravitational relaxation of clouds leading to star formation on large scale (e.g., \cite{yon97, kaw98, tac02}) and on small scale (e.g., \cite{har99, oni02}). These studies show that the virial mass derived from the velocity dispersion of molecular spectra tends to become smaller for clouds with star formation, and that star formation takes place for clouds having greater molecular column density. These are reasonable characteristics that indicate relaxation of the cloud structure under self gravity.

In order to test how the above ratio depends on star formation and cloud mass in the far-outer Galaxy, we have created a histogram of cloud mass along with the cloud mass ratios. Figure 9 clearly indicates that massive clouds having larger molecular column density form stars (dark gray) more often than lower mass ones, a similar trend to the disk clouds (see the references above). It is also seen that the virial to luminous mass ratio shown by squares and solid lines for the star-forming clouds tend to be somewhat smaller than that for non-star-forming ones (circles and broken lines). The trend seems to be consistent but not so strong as in the case of the disk clouds, possibly due to the lower spatial resolution of the present molecular data and the lower sensitivity of the infrared sources. We suggest that the present molecular clouds provide a set of good targets which can be used to make detailed searches for star formation at various wavelengths, in particular at near- to mid-infrared wavelengths.

\section{Conclusions}
We summarize the present conclusions as follows;

1) A first unbiased study of the molecular clouds in the southern Galactic Warp has detected 70 molecular clouds, nearly ten times more than previously known, in an area of 56 square degrees, thereby providing more support to the conclusion that the molecular mass in the far-outer Galaxy (we defined as $R \gtrsim$ 14.5 kpc) amounts to 2$\times$10$^7$ $M_{\odot}$.

2) The conversion factor, $N$(H$_{2}$)/$\int T(^{12}{\rm CO}) dV$, in the far-outer Galaxy derived from $L_{\rm CO}$-$M_{\rm vir}$ relation is 5.6$\times$10$^{20}$ [cm$^{-2}$/K km s$^{-1}$], 3.5($\pm1.8$) times larger than that in the inner disk. 

3) The size-linewidth relation shows offset between the inner disk and far-outer Galaxy clouds in the sense that cloud size in the far-outer Galaxy is twice that in the inner disk at any given linewidth.

4) The CO and H{\small \,I} show a good spatial and kinematical correlation although the CO clouds are much clumpier than H{\small \,I}. Warp structure and flaring are clearly seen in the molecular cloud distribution, as well as in H{\small \,I}. The total molecular mass in the far-outer Galaxy is about 1\% of that of H{\small \,I}, ten times lower than in the spiral arms and is similar to that in the interarm.

5) Only 6 of the 70 Warp clouds show signs of star formation, i.e., far-infrared compact sources detected with IRAS, although the far-infrared observations are sensitive enough only for high mass stars of B type or of earlier. The star formation efficiency derived from this is a factor of 4-5 smaller than that in the northern counterpart of the Warp.

\appendix
\section*{Comparison of CO data in the inner disk}
We have used the CO data of the inner disk clouds from the catalog of \citet{sol87} to derive the $X$ factor in the Warp (section 4). There is however a concern about the reliability of the Solomon et al.'s data because the clouds were defined at a certain high level of the temperature, i.e., 4 K \citep{sol87}. This was unavoidable due to the continuous high intensity level of the CO emission in the inner disk, even in between the clouds. On the other hand, the present Warp clouds are distributed more sparsely, allowing a cloud definition at a very low CO integrated intensity of 0.9 K km s$^{-1}$. A natural question here is if the rather high threshold value in the CO intensity in the Solomon et al.'s data may cause some errors in the cloud physical parameters in deriving the $X$ factor. It is therefore important to test if it does not cause any systematic error to use the Solomon et al.'s results in deriving the $X$ factor.

We have used the NANTEN $^{12}$CO Galactic Plan Survey results ($l <$ 60$\arcdeg$) on the Solomon et al.'s clouds. From the NANTEN dataset we have defined 55 clouds which are coincident with the distance calibrator clouds in the Solomon et al.'s catalog which includes 273 clouds, and provided three diagrams in Figure 10, i.e., $L_{\rm CO}$-$M_{\rm vir}$, size-linewidth, and $L_{\rm CO}$-linewidth relations. Details of the definition of size and linewidth are not given in BW95, while the same trend that the inner Galaxy clouds have large linewidths is clearly seen. For 11 clouds, we calculated $L_{\rm CO}$ and $M_{\rm vir}$ at three different threshold values of CO integrated intensity in order to investigate the dependence on the observational sensitivity. Spectra of the NANTEN toward No. 181 cloud (from the Solomon et al's catalog), for instance, are shown in Figure 11, for the three lowest integrated intensity (and temperature) values of 53.8(4.6), 12.8(1.3), and 5.3(0.4) K km s$^{-1}$(K), respectively. Figure 12 indicates the correlations between $L_{\rm CO}$ and $M_{\rm vir}$, showing that plots move along the least-squares fit for the 55 clouds, log($M_{\rm vir}$) = 0.91$\times$log($L_{\rm CO}$)+1.11 (c.c = 0.91).

This comparison demonstrates that the regression does not change depending on the threshold CO intensity among the NANTEN datasets and that the present analysis is fairly consistent with the Solomon et al.'s regression employed in the present work. The individual values of $L_{\rm CO}$ and $M_{\rm vir}$ of course depend on the threshold and changes a little as shown in Figure 12. Keeping in mind that the determination of the $X$ factor actually mainly depends on relative shift among the regressions in Figure 12, we conclude that the high threshold value in the Solomon et al.'s data are not likely the cause of any significant error in the present work.

\clearpage
\begin{figure}
\FigureFile(160mm,160mm){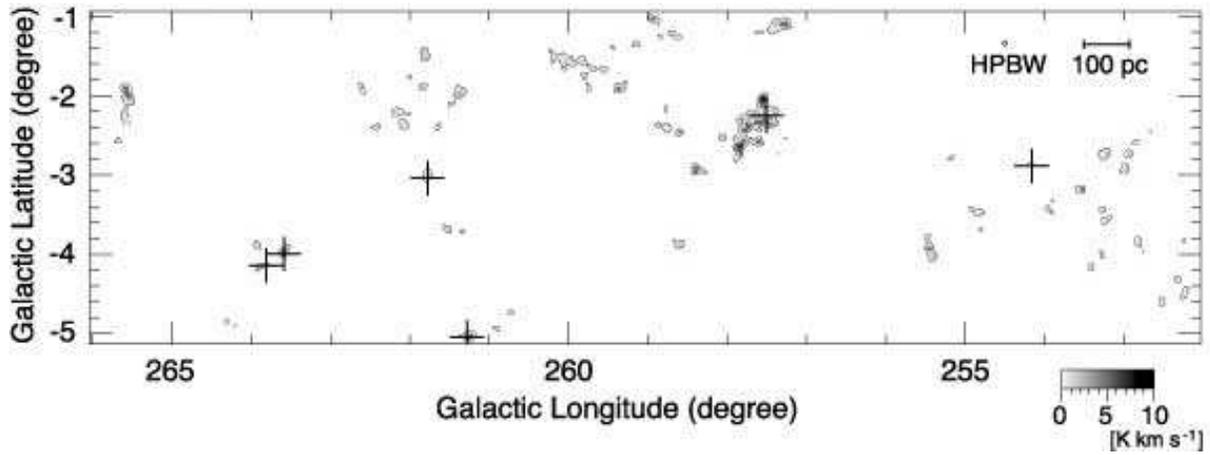}
\caption{Velocity integrated intensity distribution of $^{12}$CO ($J$ = 1--0) in the Warp region at $V_{\rm LSR}$ $\geq$ 90 km s$^{-1}$ (corresponding to $R$ = 14.9 kpc at $l$ = 252$\arcdeg$ and $R$ = 14.4 kpc at $l$ = 266$\arcdeg$, respectively). Contours are every 3.0 K km s$^{-1}$ from 0.9 K km s$^{-1}$ (3$\sigma$). Crosses indicate the position of cold IRAS sources that are associated with CO clouds.}
\end{figure}

\clearpage
\begin{figure}
\FigureFile(120mm,120mm){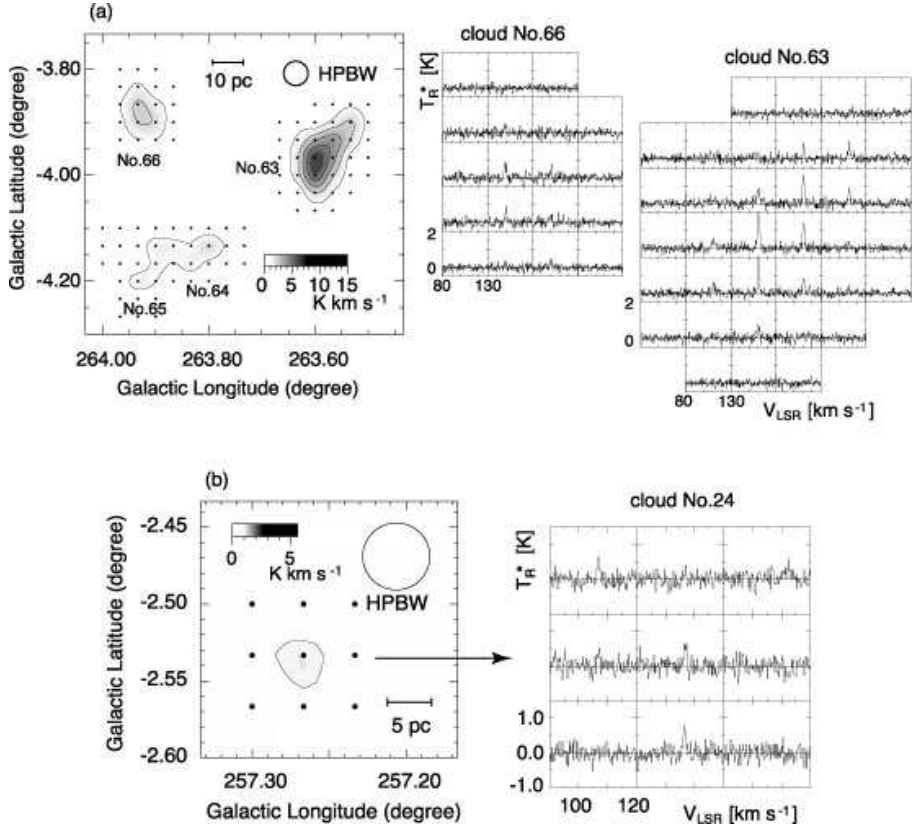}
\caption{Distribution of CO integrated intensity and CO profiles of selected Warp clouds. Contours are every 1.2 K km s$^{-1}$ from 0.9 K km s$^{-1}$. The filled circles indicate the observed positions at 2$\arcmin$ grid and cloud numbers from Table 1 are given in the individual panels. These figures show (a) typical clouds and (b) one of the smallest clouds in the present sample.}
\end{figure}

\clearpage
\begin{figure}
\FigureFile(80mm,80mm){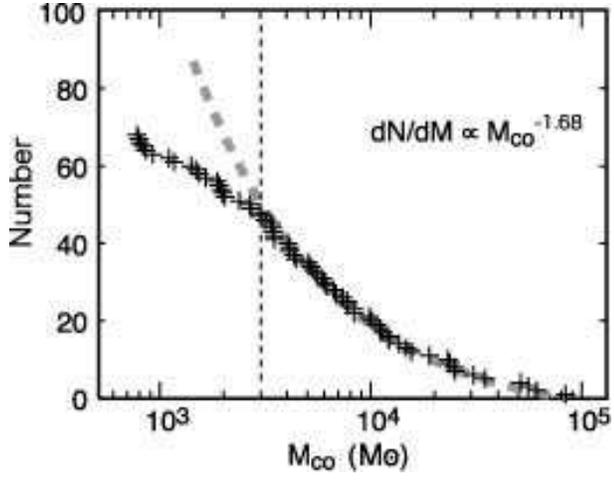}
\caption{Mass spectrum of 68 Warp clouds. The number of clouds, N($\geq M_{\rm CO}/M_{\odot}$) with mass greater than $M_{\rm CO}$, is plotted against the cloud mass $M_{\rm CO}$, along with the best-fitted power law (dashed line). The power law N($\geq M_{\rm CO}/M_{\odot}$) = 1.3$\times$10$^4$ $M_{\rm CO}^{-0.68\pm0.16}$$-$5.8 is derived by using the maximum likelihood method of \citet{cra70}.}
\end{figure}

\clearpage
\begin{figure}
\FigureFile(80mm,80mm){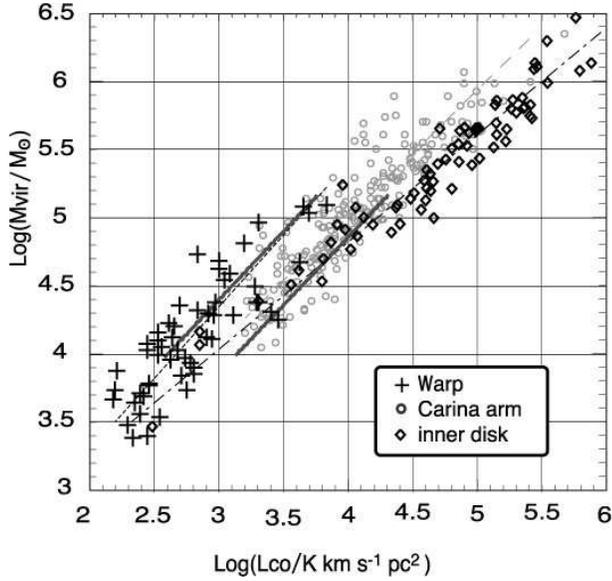}
\caption{Virial mass as a function of CO luminosity. The Warp clouds are indicated by crosses. Gray open circles show the Carina arm clouds (see Figure 8), and diamonds show the inner disk 'calibrator clouds' with known distances \citep{sol87}. Three interrupted lines are fitted by the least-squares method; Warp: log($M_{\rm vir}$) = 1.05($\pm$0.13)$\times$log($L_{\rm CO}$)+1.19($\pm$0.38), correlation coefficient(c.c.) = 0.84; Carina: log($M_{\rm vir}$) = 0.93($\pm$0.04)$\times$log($L_{\rm CO}$)+1.28($\pm$0.18), c.c. = 0.90; Inner-disk: log($M_{\rm vir}$) = 0.78($\pm$0.03)$\times$log($L_{\rm CO}$)+1.70($\pm$0.17), c.c. = 0.97.
Two thick solid lines indicate regression lines with a slope of 1.0 for the Warp and the inner disk clouds in the mass range 10$^{4.0}$-10$^{5.1}$ $M_{\odot}$; they are expressed as log($M_{\rm vir}$) = log($L_{\rm CO}$)+1.41($\pm$0.25) and log($M_{\rm vir}$) = log($L_{\rm CO}$)+0.86($\pm$0.24), respectively.}
\end{figure}

\clearpage
\begin{figure}
\FigureFile(80mm,80mm){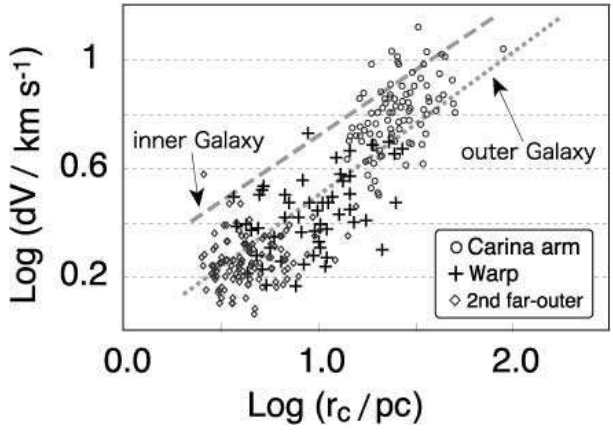}
\caption{Size-linewidth relation of the Warp clouds (crosses). For comparison, CO clouds in the Carina arm (far-side) at 9 kpc $\leq$ $R$ $\leq$ 12 kpc (circles) and from the FCRAO data with $R \geq$ 14.5 kpc by \citet{hey01} (diamonds) are shown. Broken and dotted lines in this figure for 'inner Galaxy' and 'outer Galaxy' are cited from \citet{bra95}.}
\end{figure}

\clearpage

\begin{figure}
\FigureFile(160mm,160mm){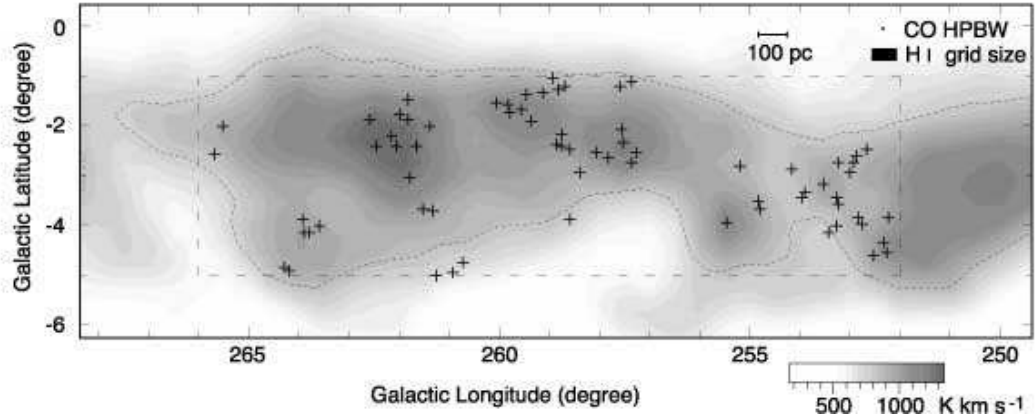}
\caption{The CO distribution is superposed on the H{\small \,I} distribution in Galactic coordinates. The gray-scale map indicates the H{\small \,I} velocity integrated intensity of 90-140 km s$^{-1}$ in $V_{\rm LSR}$ (approximately corresponding to $R$ $\sim$14.5-25 kpc), and thin dotted contour indicates the integrated intensity = 800 K km s$^{-1}$ contour level of H{\small \,I}. Crosses indicate the positions of CO clouds. The broken line shows the observed area.}
\end{figure}

\clearpage
\begin{figure}
\FigureFile(80mm,80mm){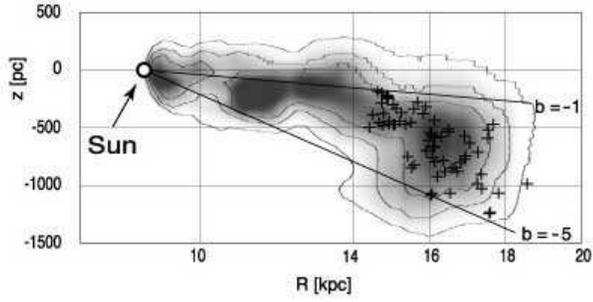}
\caption{The Warp clouds are shown in a plane; the distance from the Galactic center ($R$) vs. distance from the plane of $b$ = 0$\arcdeg$ ($z$). Crosses indicate the peak positions of CO clouds in the Warp ($V_{\rm LSR}$ $\geq$ 90 km s$^{-1}$) and the gray-scale indicate H{\small \,I} distribution integrated over Galactic longitude from 252$\arcdeg$ to 266$\arcdeg$. Contours show 30\%, 45\% and 60\% levels of the maximum H{\small \,I} intensity (located at $R$ = 11.5 kpc and $z$ = $-$250 pc). The observed area is denoted by solid lines.}
\end{figure}

\clearpage
\begin{figure}
\FigureFile(160mm,160mm){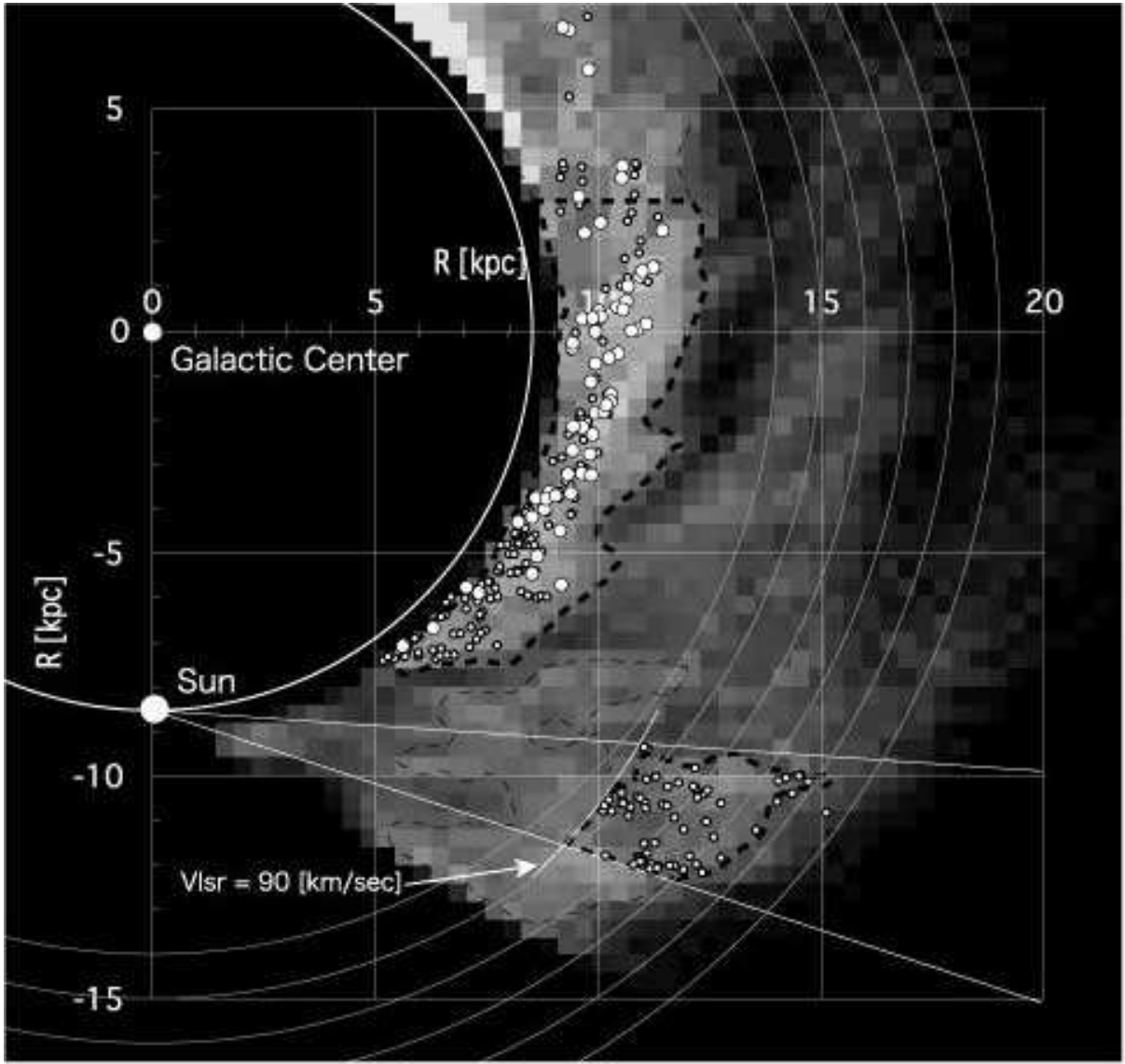}
\caption{Face-on view of H{\small \,I} (gray scale) in the outer galaxy and CO peak positions in the Carina arm and the Warp region. Larger circles in the Carina arm show molecular clouds with $M_{\rm CO} >$ 10$^5$ $M_{\odot}$.}
\end{figure}

\clearpage
\begin{figure}
\FigureFile(80mm,80mm){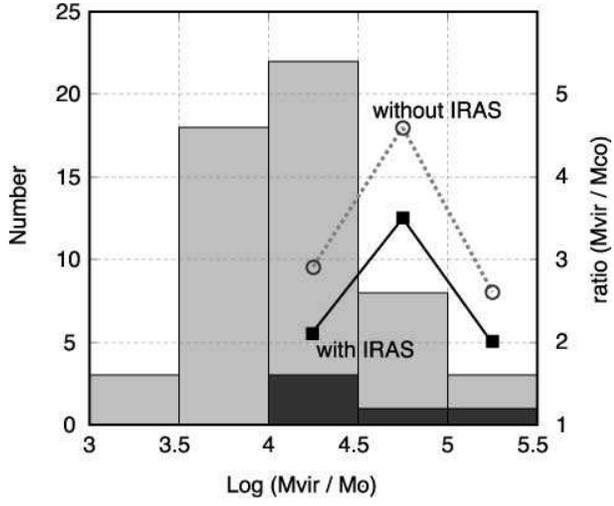}
\caption{$M_{\rm vir}$ histogram for the Warp clouds with (dark gray) and without (light gray) IRAS sources (left axis). Averaged $M_{\rm vir}$/$M_{\rm CO}$ ratio for clouds with (squares and solid lines) and without IRAS (circles and dotted lines) in each bin are presented at the right axis.}
\end{figure}

\clearpage
\begin{figure}
\FigureFile(160mm,160mm){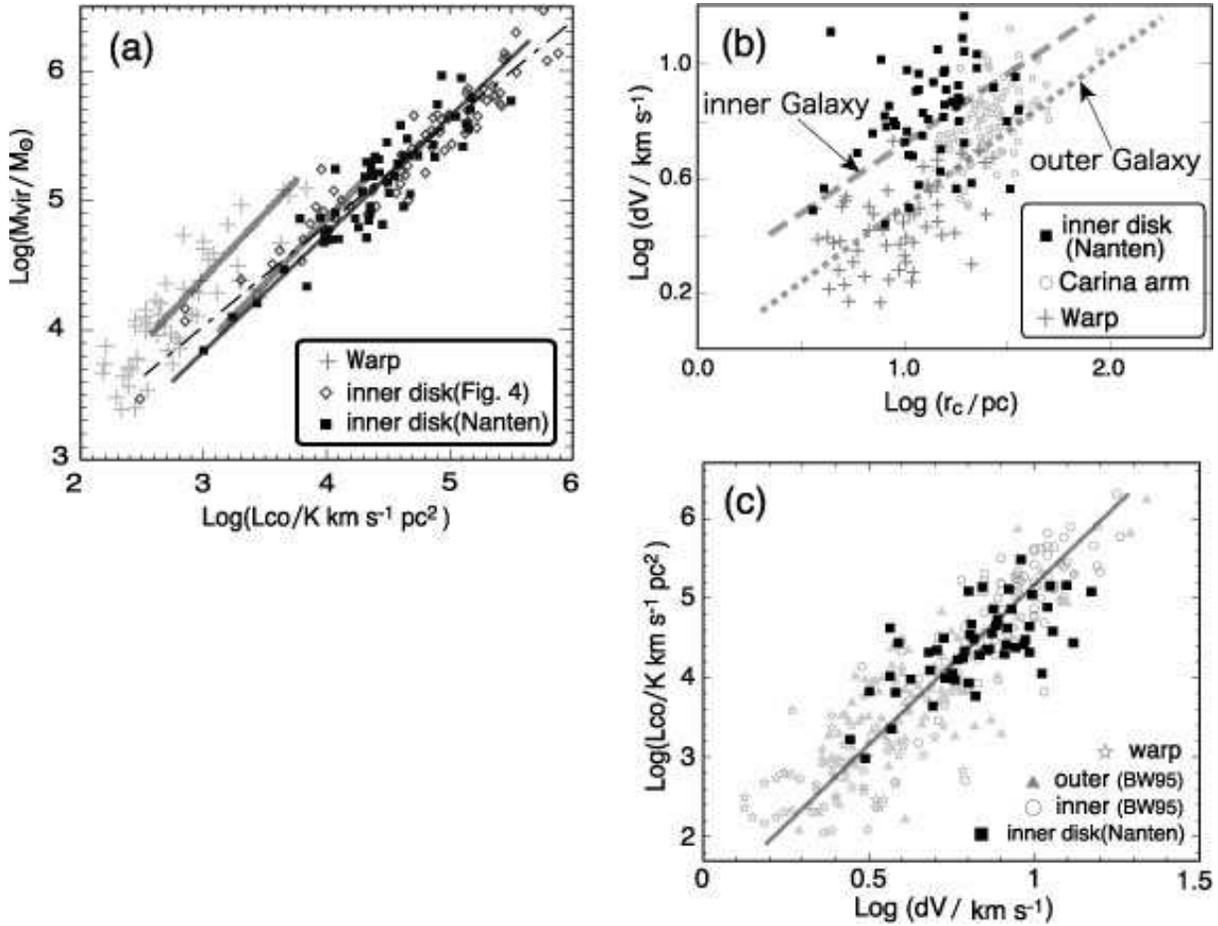}
\caption{(a)$L_{\rm CO}$-$M_{\rm vir}$: 55 inner clouds identified by the NANTEN data, 75 inner clouds identified by Solomon et al. (broken line) and the Warp clouds are plotted. The long solid line shows the least-squares fit for 55 clouds, and the parallel lines are same as that in Figure 4. (b)Size-linewidth: 55 clouds are plotted. Other plots and lines are same as Figure 5. Our data is consistent with BW95's result, with the tendency of larger linewidth of inner disk clouds. (c)linewidth-$L_{\rm CO}$: 55 clouds, the Warp clouds, the inner and outer clouds from BW95 are plotted. This figure shows a good agreement with the empirical relation that $L_{\rm CO}$ is proportional to linewidth to the 4th power (thick line)} 
\end{figure}

\clearpage
\begin{figure}
\FigureFile(80mm,80mm){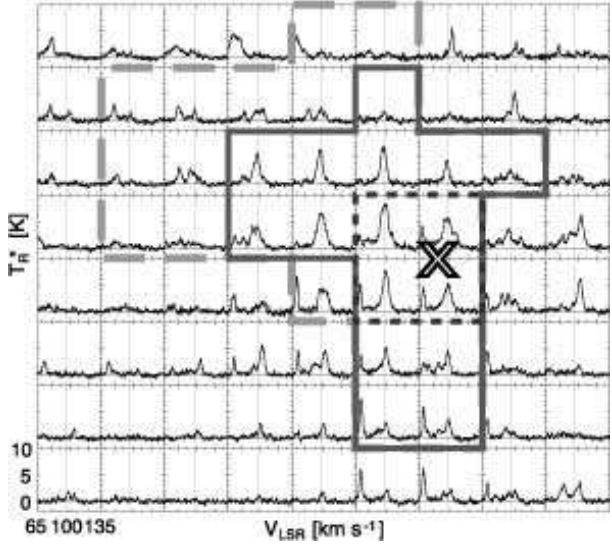}
\caption{Spectra from the NANTEN data toward No. 181 cloud, which number is cited from Table 1 of \citet{sol87}. $X$ mark represents the center position which \citet{sol87} gave. Dotted lines show the center velocity, $V_{\rm LSR}$ = 97 km s$^{-1}$. We have originally defined this cloud with 16 points surrounded by solid line. Within the dashed line, more 11 spectra are included to the left-top direction. Those 11 extra spectra, having irregular shapes in the same $V_{\rm LSR}$ component toward the peak position, may be part of this cloud. The dotted line cuts out only 4 spectra toward the central part. Then we have also calculated $L_{\rm CO}$ and $M_{\rm vir}$ for both larger and smaller clouds.}
\end{figure}

\clearpage
\begin{figure}
\FigureFile(80mm,80mm){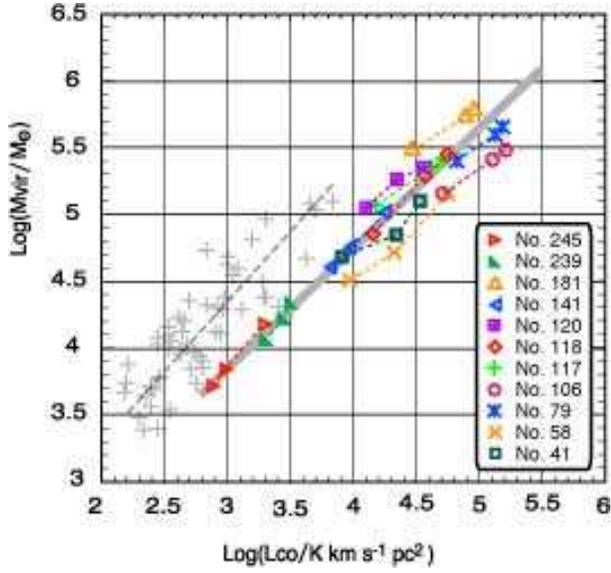}
\caption{For eleven clouds, we have calculated $L_{\rm CO}$ and $M_{\rm vir}$ at three different threshold values of CO integrated intensity in a same manner in Figure 11, and all 33 points are plotted on the $L_{\rm CO}$-$M_{\rm vir}$ diagram, compared with the Warp clouds (crosses), same as Figure 4. Thick line shows the least-squares fit for 55 inner clouds identified by the NANTEN data.}
\end{figure}

\end{document}